\documentclass{aa}
\usepackage{graphicx,natbib}
\bibpunct{(}{)}{;}{a}{}{,}

\newcommand{\lx}{L_\mathrm{x}}
\newcommand{\lxs}{L_\mathrm{s}}
\newcommand{\lxh}{L_\mathrm{h}}
\newcommand{\ms}{M_\odot}
\newcommand{\vm}{V_\mathrm{max}}
\newcommand{\vmm}{V_\mathrm{gen}}

\newcommand{\lesssim}{\mathrel{\hbox{\rlap{\hbox{\lower4pt\hbox{$\sim$}}}\hbox{$<$}}}}
\newcommand{\gtrsim}{\mathrel{\hbox{\rlap{\hbox{\lower4pt\hbox{$\sim$}}}\hbox{$>$}}}}
\newcommand{\beq}{\begin{equation}}
\newcommand{\eeq}{\end{equation}}
\newcommand{\beqa}{\begin{eqnarray}}
\newcommand{\eeqa}{\end{eqnarray}}

\begin{document}

\title{X-ray luminosity function of faint point sources in the
Milky Way}   

\author{S.~Sazonov\inst{1,2} \and M.~Revnivtsev\inst{1,2} \and
  M.~Gilfanov\inst{1,2} \and E.~Churazov~\inst{1,2} \and R.~Sunyaev\inst{1,2}}

\offprints{sazonov@mpa-garching.mpg.de}

\institute{Max-Planck-Institut f\"ur Astrophysik,
           Karl-Schwarzschild-Str. 1, D-85740 Garching bei M\"unchen,
           Germany
     \and   
           Space Research Institute, Russian Academy of Sciences,
           Profsoyuznaya 84/32, 117997 Moscow, Russia
}
\date{Received / Accepted}

\authorrunning{Sazonov et al.}
\titlerunning{X-ray luminosity function}

\abstract{We assess the contribution to the X-ray (above 2~keV) luminosity 
of the Milky Way from different classes of low-mass binary systems and
single stars. We begin by using the RXTE Slew Survey of the sky at 
$|b|>10^\circ$ to construct an X-ray luminosity function (XLF) of
nearby X-ray sources in the range $10^{30}~{\rm erg}~{\rm
s}^{-1}<\lx<10^{34}~{\rm erg}~{\rm s}^{-1}$ (where $\lx$ is the
luminosity over 2--10~keV), occupied by coronally active binaries
(ABs) and cataclysmic variables (CVs). We then extend this XLF down to
$\lx\sim 10^{27.5}~{\rm erg}~{\rm s}^{-1}$ using the Rosat All-Sky
Survey in soft X-rays and available information on the 0.1--10~keV
spectra of typical sources. We find that the local cumulative X-ray
(2--10~keV) emissivities (per unit stellar mass) of ABs and CVs are
$(2.0\pm 0.8)\times 10^{27}$ and $(1.1\pm 0.3)\times
10^{27}$~erg~s$^{-1}$~$\ms^{-1}$, respectively. In addition to ABs and
CVs, representing old stellar populations, young stars emit locally
$(1.5\pm 0.4)\times 10^{27}$~erg~s$^{-1}\ms^{-1}$. We finally attach
to the XLF of ABs and CVs a high luminosity branch (up to $\sim
10^{39}$~erg~s$^{-1}$) composed of neutron-star and black-hole
low-mass X-ray binaries (LMXBs), derived in previous work. The
combined XLF covers $\sim 12$ orders of magnitude in luminosity. The
estimated combined contribution of ABs and CVs to the 2--10~keV
luminosity of the Milky Way is $\sim 2\times 10^{38}$~erg~s$^{-1}$,
or $\sim$3\% of the integral luminosity of LMXBs (averaged over nearby
galaxies). The XLF obtained in this work is used elsewhere (Revnivtsev et
al.) to assess the contribution of point sources to the Galactic ridge
X-ray emission.

}
\maketitle

\section{Introduction}
\label{intro}

X-ray (above 2~keV) emission is a ubiquitous property of
different classes of low-mass close binaries, ranging in order of
increasing luminosity from chromospherically and coronally active 
binaries (ABs) through cataclysmic variables (CVs, magnetic 
and non-magnetic) and related white-dwarf accretors (symbiotic stars) to 
neutron-star and black-hole binaries (LMXBs). Although each of these
classes has been thoroughly investigated for decades, there remains  
significant uncertainty as regards the contribution of ABs and CVs,
both cumulative and as a function of luminosity, to the integral X-ray
luminosity of the Galaxy. On the other hand, the luminosity function
(XLF) of LMXBs has been measured with good precision for the Milky Way
and nearby galaxies \citep{ggs02,gilfanov04}. 

There are several important astrophysical problems urging a
detailed study of the XLF of ABs and CVs. First, there is
a long-standing puzzle of the origin of the Galactic ridge X-ray
emission (e.g. \citealt{wmb+82,wtw+85}). Although there were early
suggestions that this apparently diffuse X-ray emission might be
composed of thousands and millions of CVs and ABs
\citep{wm83,os92,ms93}, recent deep surveys by Chandra and XMM-Newton
in the Galactic plane and in the Galactic Center region resolved only
$\sim$10--30\% of the ridge emission (above 2~keV) into point sources
\citep{mbb+04,etp+05,hww+04}, leaving open the question as to what
fraction of the unresolved emission is truly diffuse. 

Secondly, with the advent of Chandra it has become possible to
obtain high-quality X-ray maps of nearby elliptical galaxies and
resolve on them individual LMXBs. The underlying diffuse
emission is usually attributed to the hot ($\sim 0.5$~keV)
interstellar gas. However, in gas poor galaxies
unresolved point X-ray sources associated with the old stellar
population may contribute significantly to or even dominate the
apparently diffuse emission, especially at high energies
(e.g. \citealt{cft87,mka+97,iab03}). The interpretation of X-ray
observations of gas poor ellipticals thus depends critically on our
knowledge of the XLF of low-mass binaries in these galaxies, which is
expected to resemble the XLF of ABs, CVs and LMXBs in our Galaxy
scaled by the stellar mass. 

With the above motivation in mind, we construct below a combined XLF
of ABs, CVs and LMXBs covering the very broad luminosity range from
$\lx\sim 10^{27.5}$ to $\sim 10^{39}$~erg~s$^{-1}$ (where $\lx$ is the
luminosity in the 2--10~keV band). Since these classes of objects
represent  old stellar populations (in particular ABs maintain high
levels of activity throughout their lives due to tidal locking of rapid
stellar rotation) their XLF normalized to the stellar mass is not
expected to vary significantly across the Galaxy and also between
different types of galaxies. This has already been observationally
demonstrated for LMXBs \citep{gilfanov04}. In contrast, the statistics of
young coronal stars (YSs), another abundant class
of low-luminosity X-ray sources (see \citealt{g04} for a review), is
expected to be governed by local star formation history, so that an XLF
constructed for this class of sources in the solar neighborhood may
not be representative of other parts of the Galaxy and other
galaxies. It will be shown below that in the solar neighborhood YSs
produce $\sim 30$\% of the integral 2--10~keV emissivity.

Our assessment of source space densities at  
$\lx<10^{34}$~erg~s$^{-1}$ will be based on the RXTE Slew Survey
(\citealt{rsj+04}, hereafter R04) and Rosat All-Sky Survey
(http://www.xray.mpe.mpg.de/cgi-bin/rosat/rosat-survey; \citealt{vab+99}).
In the latter case we also employ spectral information from various 
X-ray missions to convert the derived XLF from a soft X-ray band to the
standard X-ray band. The high-luminosity ($\lx>10^{34}$~erg~s$^{-1}$) 
branch of the XLF is adopted from previous work of \cite{gilfanov04}. 

\section{Medium luminosity range: RXTE Slew Survey}
\label{xss}

Recently, a serendipitous survey of the whole sky  in the
3--20~keV energy band was performed based on slew observations with
the PCA instrument on the RXTE spacecraft (RXTE Slew Survey, or XSS),
and a source catalog at high Galactic latitude ($|b|>10^\circ$) was
produced (R04). The survey achieved a flux limit of $2.5\times
10^{-11}$~erg~cm$^{-2}$~s$^{-1}$ (3--20~keV) or better for 90\% of the
$|b|>10^\circ$ sky.

The majority of the 294 detected XSS sources have been identified with
extragalactic objects. 60 sources have been identified with objects in
the Galaxy, while 21 sources still remain unidentified. The 
identified Galactic sample includes 14 LMXBs and HMXBs, which will not
be considered below. We also exclude from the current consideration 4
star forming complexes with multiple X-ray sources as unresolvable by 
RXTE (Orion, Chamaeleon 1, Chamaeleon 2 and $\rho$ Ophiuchi), the hot
supergiant star $\zeta$ Ori as also belonging to the Orion complex,
the unresolved globular cluster NGC 6397, and supernova remnant SN 1006
as an extended X-ray source. This leaves us with a total of 40
identified ABs (including 2 candidates, see below) and CVs. Of these we 
additionally excluded 10 sources for either of the following reasons:
1) the source is not detectable on the average XSS map and was
originally included in the XSS catalog based on its transient
detection, 2) the source was the target of pointed RXTE observations
and would not have been detected in slew observations otherwise.


\begin{table*}
\caption{XSS sources identified with ABs and CVs 
\label{xss_sources}
}
\smallskip

\begin{center}

\begin{tabular}{lllrrlll}
\hline
\hline
\multicolumn{1}{c}{XSS source} &
\multicolumn{1}{c}{Name} &
\multicolumn{1}{c}{Class$^{\rm a}$} & 
\multicolumn{1}{c}{$D^{\rm b}$} & 
\multicolumn{1}{c}{$\lxh^{\rm c}$} &
\multicolumn{1}{c}{$\lx^{\rm d}$} &
\multicolumn{1}{l}{Hardness} &
\multicolumn{1}{l}{$1/V_\mathrm{gen}$}
\\   
\multicolumn{1}{c}{(J2000.0)} &
\multicolumn{1}{c}{} &     
\multicolumn{1}{c}{} & 
\multicolumn{1}{c}{pc} & 
\multicolumn{1}{c}{erg/s} &   
\multicolumn{1}{c}{erg/s} &   
\multicolumn{1}{c}{ratio} &
\multicolumn{1}{c}{pc$^{-3}$} 
\\     
\hline
02290$-$6931 & RBS 324      & P        &  250$^{~1}$ & 31.88 & 31.70 & 0.41$\pm$0.25 & $4.9\times 10^{-8}$ \\
02569+1931   & XY Ari       & IP       &  270$^{~2}$ & 32.43 & 32.12 & 0.88$\pm$0.42 & $1.2\times 10^{-8}$ \\
03089+4101   & $\beta$ Per  & AL       &   28$^{~3}$ & 30.91 & 30.77 & 0.28$\pm$0.06 & $8.9\times 10^{-7}$ \\
03385+0029   & V711 Tau     & RS       &   29$^{~3}$ & 30.23 & 30.15 & 0.10$\pm$0.19 & $8.2\times 10^{-6}$ \\ 
05019+2444   & V1062 Tau    & IP       & 1100$^{~4}$ & 33.77 & 33.51 & 0.69$\pm$0.12 & $1.2\times 10^{-9}$ \\
05295$-$3252 & TV Col       & IP       &  370$^{~5}$ & 33.08 & 32.80 & 0.77$\pm$0.09 & $3.1\times 10^{-9}$ \\
05432$-$4116 & TX Col       & IP       &  500$^{~4}$ & 32.71 & 32.50 & 0.52$\pm$0.21 & $6.3\times 10^{-9}$ \\
05450+6049   & BY Cam       & P        &  190$^{~6}$ & 31.96 & 31.74 & 0.55$\pm$0.24 & $3.9\times 10^{-8}$ \\
06132+4755   & SS Aur       & DN       &  280$^{~7}$ & 32.17 & 31.97 & 0.48$\pm$0.24 & $2.2\times 10^{-8}$ \\  
07514+1442   & PQ Gem       & IP       &  400$^{~4}$ & 32.79 & 32.54 & 0.68$\pm$0.20 & $5.3\times 10^{-9}$ \\
08010+6241   & HT Cam       & IP       &  400$^{~8}$ & 32.72 & 32.46 & 0.70$\pm$0.29 & $6.2\times 10^{-9}$ \\
08142+6231   & SU Uma       & DN       &  260$^{~7}$ & 32.16 & 31.99 & 0.37$\pm$0.17 & $2.3\times 10^{-8}$ \\
11474+7143   & DO Dra       & IP       &  155$^{~4}$ & 32.02 & 31.78 & 0.62$\pm$0.10 & $3.3\times 10^{-8}$ \\
12392$-$3820 & V1025 Cen    & IP       &  400$^{~9}$ & 32.36 & 32.19 & 0.37$\pm$0.14 & $1.4\times 10^{-8}$ \\
12529$-$2911 & EX Hya       & IP       &   65$^{10}$ & 31.79 & 31.62 & 0.40$\pm$0.02 & $6.2\times 10^{-8}$ \\
13355+3714   & BH CVn       & RS       &   45$^{~3}$ & 31.15 & 31.07 & 0.11$\pm$0.11 & $4.2\times 10^{-7}$ \\
14100$-$4500 & V834 Cen     & P        &  150$^{11}$ & 31.72 & 31.48 & 0.61$\pm$0.12 & $7.6\times 10^{-8}$ \\
14241$-$4803 & HD125599$^{\rm e}$ & ?  &   90$^{~3}$ & 30.72 & 30.62 & 0.15$\pm$0.15 & $1.6\times 10^{-6}$ \\
14527$-$2414 & HD130693$^{\rm f}$ & ?  &   27$^{12}$ & 30.07 & 29.93 & 0.28$\pm$0.08 & $1.4\times 10^{-5}$ \\
16167$-$2817 & V893 Sco     & DN       &  150$^{~7}$ & 32.10 & 31.92 & 0.40$\pm$0.08 & $2.7\times 10^{-8}$ \\
17309$-$0552 & 1RXS J173021.5-055933 &IP&3300$^{13}$ & 34.28 & 34.06 & 0.54$\pm$0.29 & $8.1\times 10^{-10}$\\ 
17597+0821   & V2301 Oph    & P        &  150$^{14}$ & 31.90 & 31.67 & 0.56$\pm$0.10 & $4.6\times 10^{-8}$ \\
18080+0622   & V426 Oph     & DN       &  200$^{15}$ & 32.07 & 31.85 & 0.56$\pm$0.25 & $2.9\times 10^{-8}$ \\
18164+5004   & AM Her       & P        &   79$^{~7}$ & 31.77 & 31.50 & 0.74$\pm$0.05 & $6.6\times 10^{-8}$ \\
18553$-$3111 & V1223 Sgr    & IP       &  510$^{16}$ & 33.73 & 33.45 & 0.76$\pm$0.02 & $1.2\times 10^{-9}$ \\
19243+5041   & CH Cyg       & SS       &  250$^{17}$ & 32.26 & 31.87 & 1.25$\pm$0.26 & $1.8\times 10^{-8}$ \\
21155$-$5836 & CD Ind       & P        &  250$^{18}$ & 31.90 & 31.64 & 0.70$\pm$0.24 & $4.6\times 10^{-8}$ \\
22178$-$0822 & FO Aqr       & IP       &  300$^{~4}$ & 32.85 & 32.53 & 0.94$\pm$0.09 & $4.7\times 10^{-9}$ \\
22526+1650   & IM Peg       & RS       &   97$^{~3}$ & 31.65 & 31.46 & 0.42$\pm$0.03 & $9.3\times 10^{-8}$ \\
22551$-$0309 & AO Psc       & IP       &  250$^{~4}$ & 32.57 & 32.33 & 0.60$\pm$0.10 & $8.6\times 10^{-9}$ \\
\hline             
\end{tabular}

\end{center}

$^{\rm a}$ Class: RS -- RS CVn, AL -- Algol, DN -- dwarf nova, P --
polar, IP -- intermediate polar, SS -- symbiotic star

$^{\rm b}$ Reference for distance: 1 -- \cite{sbb+02}, 2 --
\cite{ldm01}, 3 -- Hipparcos, 4 -- \cite{p94}, 5 -- \cite{mbl+01}, 6
-- \cite{w95}, 7 -- \cite{t03}, 8 -- \cite{tgk+98}, 9 --
assumed, 10 -- \cite{ebr+02}, 11 -- \cite{agl+05}, 12 -- Tycho, 13 --
based on \cite{gme+05}, see main text, 14 -- \cite{srh+94}, 15 --
\cite{h98}, 16 -- \cite{bhm+04}, 17 -- \cite{sk03}, 18 -- lower limit
(The MSSL Polar Page, http://www.mssl.ucl.ac.uk/www\_astro/gal/polar.html)  
 
$^{\rm c}$ Log of 3--20~keV luminosity

$^{\rm d}$ Log of 2--10~keV luminosity

$^{\rm e}$ Bright (V=8.5) F7/8V star associated with the bright ROSAT
source 1RXS J142148.7$-$480420 

$^{\rm f}$ Bright (V=8.2) G6V star associated with the bright ROSAT
source 1RXS J145017.6$-$242558=RBS
1436 \citep{shl+00}

\end{table*}


We have thus obtained a sample (see Table~\ref{xss_sources}) of 30 ABs and CVs
detected with $\ge 4\sigma$ significance on the average XSS sky map
(3--20 keV). This sample is well suited for statistical studies. 

For each source, the XSS catalog provides RXTE/PCA count rates in
the 3--8~keV and 8--20~keV bands. We found published distances
to all sources except for the intermediate polar V1025 Cen (for which
we assumed a distance of 400~pc, a value typical for intermediate
polars in our sample), the polar CD Ind (for which we used the
available lower limit) and the source XSS J17309$-$0552 discussed
below. Parallax measurements, in many cases adopted 
directly from the Hipparcos or Tycho catalog, were used wherever
available. For XSS J17309$-$0552/RXS J173021.5$-$055933, a
recently discovered intermediate polar \citep{gme+05}, we estimated
the distance using available information about the secondary star
\citep{gme+05}. Specifically, this GV star contributes $\sim
15$\% to the $R$-band flux of the binary. Given the
system's visual magnitude ($R_V=15.4$) and interstellar extinction 
toward it [$E(B-V)\sim 0.45$], and assuming that the secondary is on
the main sequence, we find a distance $\sim$2300--3100~pc. However,
the very long orbital period of the binary (15.4 hours) implies that
the Roche-lobe filling secondary is evolved
(e.g. \citealt{sd98}). This yields a more likely distance of 
$\sim 3,300$~pc, which we adopt here. 

 Using the distance estimates and measured source count
rates in the 3--8~keV and 3--20~keV bands, we determined source
luminosities in the 2--10~keV ($\lx$)  and 3--20~keV ($\lxh$) band,
respectively. A Crab-like spectrum was assumed for this calculation,
which is expected to ensure reasonable accuracy of energy flux
estimation for our sources given their measured hardness
ratios (8--20~keV counts over 3--8~keV counts). We note that the quoted
luminosities are observed ones, i.e. they are not corrected for any
absorption intrinsic to the sources. The interstellar absorption toward our
(high Galactic latitude) sources is not expected to have a significant
effect on the RXTE measured fluxes. This is true even in the case 
of XY Ari, the only source in our sample known to be located behind a molecular
cloud, for which we estimate a line-of-sight absorption of
$N_{\rm H}\sim 2\times 10^{22}$~cm$^{-2}$ from the measured visual
extinction $A_{V}\sim 11.5$ \citep{ldm01}. Similarly the 
uncertainty in source distances is unlikely to significantly affect
the statistical results obtained below.  

\begin{figure}
\centering
\includegraphics[width=\columnwidth]{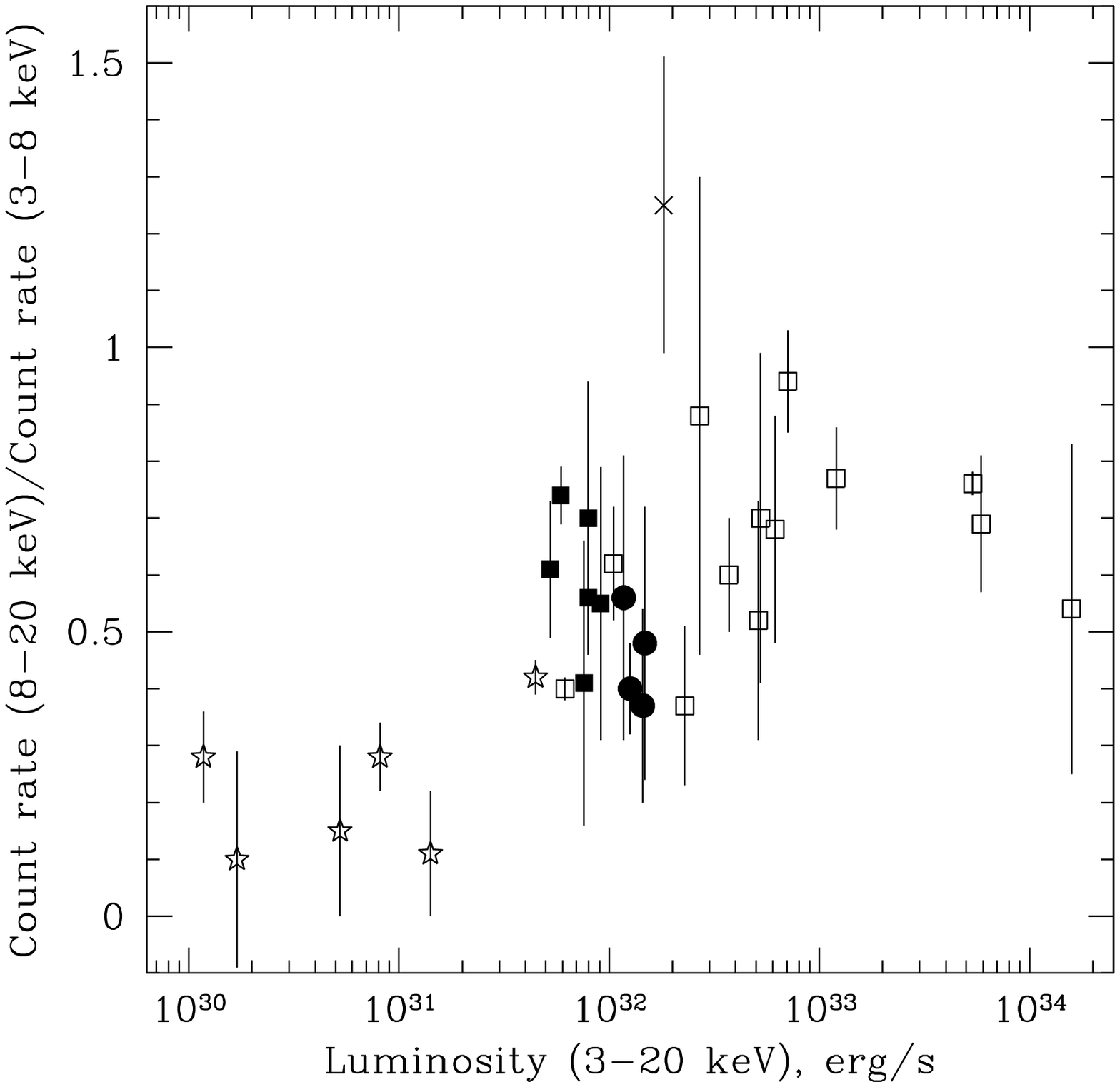}
\caption{RXTE/PCA hardness ratio vs. 3--20~keV luminosity for XSS
sources identified 
with ABs and CVs: open stars -- ABs, filled circles -- dwarf novae,
filled squares -- polars, open squares -- intermediate 
polars, the cross -- a symbiotic star.
} 
\label{hardness_lum}
\end{figure}

Our XSS sample includes 6 known or suspected ABs and 24 known
CVs. Among the former there are 3 RS CVn binaries, the
prototype Algol system ($\beta$ Per) and 2 late-type main-sequence
stars (HD125599 and HD130693) which we consider candidate ABs based on
their optical spectral class, X-ray luminosities and relative X-ray
softness compared to CVs. The CV subsample includes 4 non-magnetic CVs
(dwarf novae), 19 magnetic CVs (6 polars and 13 intermediate polars)
and 1 symbiotic star.

In Fig.~\ref{hardness_lum} we plot the XSS hardness ratio  as a
function of luminosity for our identified sample. One can see 
that, as expected, the ABs have softer spectra than the
CVs. As regards the latter, the intermediate polars and the only
symbiotic star in our sample tend to have somewhat harder spectra than
the polars and dwarf novae.

Despite the relatively small size of our sample, it can be used for
construction of an XLF since it covers 4 orders of magnitude in
luminosity (from $\sim 10^{30}$ to $\sim 10^{34}$~erg~s$^{-1}$) and is derived
from a flux limited survey (with a direction dependent sensitivity
limit). As usual for such surveys, one can readily estimate the space 
density of sources in a given luminosity interval using the $1/\vm$
method \citep{s68}. To this end, we use the XSS exposure map presented
in R04, which yields the space volume probed by the survey for a given
source luminosity.  

To take into account the fact that the studied classes of sources
are concentrated toward the Galactic plane, we assume that the space
density of ABs and CVs declines with height as $\exp (-z/h)$, where
$h=150$~pc. This adopted scale height is appropriate for the CVs
(e.g. \citealt{p84}), while the inferred space 
density of ABs is only weakly sensitive to the assumed value of $h$
(since ABs are detectable within $\sim$100~pc of the Sun in the XSS). Note that
the Galactocentric dependence is not important for us since we study
objects  within $\sim 1$~kpc of the Sun. We therefore weight the
standard $\delta\vm$ volume found for each small solid angle
$\delta\Omega$ (at Galactic latitude $b$) of the survey by the
space density of sources integrated over $\delta\Omega$ and over
distance from 0 to $d_{\rm max}$, the maximum distance at which a given XSS
source is detectable \citep{trm93,sbb+02}: 
\beq
\delta\vmm=\delta\Omega\frac{h^3}
{\sin^3 b}\left[2-(\xi^2+2\xi+2)e^{-\xi}\right], 
\eeq
where $\xi=d_{\rm max}\sin b/h$. Each
sampled source then contributes $1/\sum\delta\vmm$ to the estimated
space density and $1/(\sum\delta\vmm)^2$ to the associated variance,
where the sum is taken over the total solid angle of the survey. 

\begin{figure}
\centering
\includegraphics[width=\columnwidth]{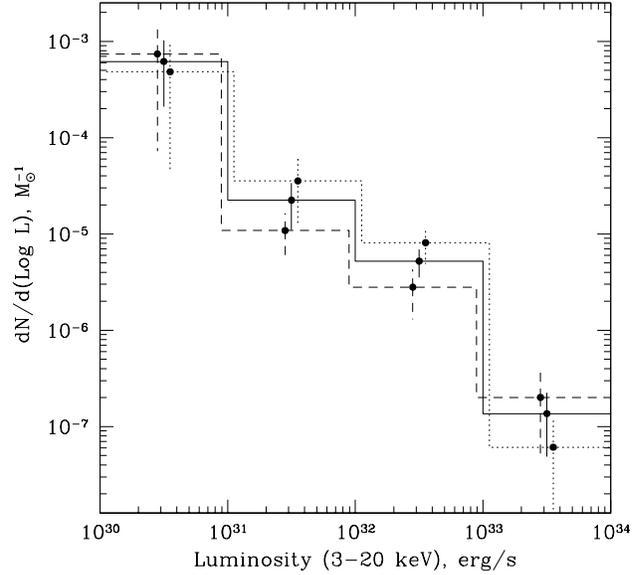}
\caption{Differential 3--20~keV luminosity function of ABs and CVs
derived from the XSS. Solid, dotted and dashed histograms and
error bars show the XLFs for the whole, northern and southern 
sky, respectively. The northern and southern XLFs are slightly shifted 
along the luminosity axis for better visibility.  
} 
\label{lumfunc_rxte}
\end{figure}

We show in Fig.~\ref{lumfunc_rxte} the resulting differential XLF of
ABs and CVs in the 3--20~keV energy band, covering the luminosity range
$10^{30}$--$10^{34}$~erg~s$^{-1}$. This XLF was normalized to the local
stellar mass density, assumed to be $0.04\ms$~pc$^{-3}$ throughout the
paper \citep{jw97,rrd+03}. The values of $1/\vmm$ for individual XSS
sources are given in Table~\ref{xss_sources}. Note that we
excluded the intermediate polar XSS J17309$-$0552/RXS
J173021.5$-$055933 from the XLF construction since its inferred X-ray
luminosity exceeds $10^{34}$~erg~s$^{-1}$ making it the only
source with such high luminosity in our sample.

It is necessary to check  whether the derived XLF suffers from incompleteness
of the input sample. There are in fact 18 unidentified XSS
sources\footnote{Note that \cite{sr04} listed 35  
unidentified XSS sources, but 5 of those sources were transiently
detected and another 12 have been identified since publication} --
see Table~\ref{xss_noids}. Although, we suspect (see R04) that
most of these sources are active galactic nuclei, this has not yet
been verified and therefore we must take this additional sample into
account. 


\begin{table}
\caption{Unidentified XSS sources
\label{xss_noids}
}
\smallskip

\begin{tabular}{lll}
\hline
\hline
\multicolumn{1}{c}{XSS source} &
\multicolumn{1}{c}{Hardness} &
\multicolumn{1}{c}{Counterpart from RASS}
\\   
\multicolumn{1}{c}{(J2000.0)} &
\multicolumn{1}{c}{ratio} &
\multicolumn{1}{c}{Bright Source Catalog}
\\     
\hline
00050$-$6904 &  $0.46\pm0.11$ & \\ 
00564+4548   &  $0.52\pm0.08$ & 1RXS J005528.0+461143\\
02087$-$7418 &  $0.86\pm0.23$ & \\       
05188+1823   &  $0.70\pm0.34$ & \\                      
12270$-$4859 &  $0.52\pm0.13$ & 1RXS J122758.8$-$485343\\   
13563$-$7342 &  $0.46\pm0.26$ & \\
14101$-$2936 &  $1.19\pm0.47$ & \\   
14138$-$4022 &  $0.46\pm0.23$ & \\ 
14239$-$3800 &  $0.51\pm0.21$ & 1RXS J142149.8$-$380901\\
14353$-$3557 &  $0.59\pm0.33$ & \\ 
14495$-$4005 &  $0.45\pm0.15$ & \\ 
15360$-$4118 &  $0.63\pm0.29$ & \\
16049$-$7302 &  $0.76\pm0.32$ & \\ 
16537$-$1905 &  $0.70\pm0.24$ & \\
17223$-$7301 &  $0.37\pm0.27$ & 1RXS J171850.0$-$732527$^{\rm a}$\\
17576$-$4534 &  $0.63\pm0.30$ & \\
18486$-$2649 &  $0.43\pm0.24$ & \\
19303$-$7950 &  $0.49\pm0.24$ & 1RXS J194944.6$-$794519\\
\hline             
\end{tabular}

$^{\rm a}$ Possibly associated with star Tyc 9288-744-1 (V=9.8)  

\end{table}

\begin{figure}
\centering
\includegraphics[width=\columnwidth]{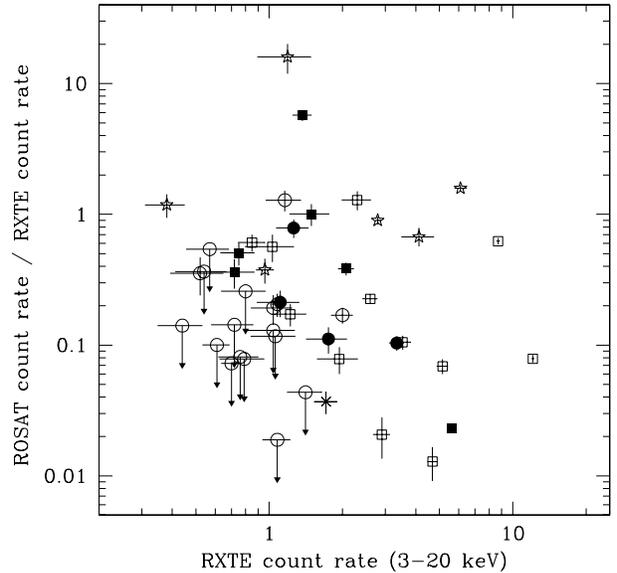}
\caption{Ratio of the ROSAT/PSPC count rate (0.1--2.4~keV) to the
RXTE/PCA count rate (3--20~keV) for the identified and unidentified
XSS sources. Identified sources of different classes are denoted by
the same symbols as in Fig.~\ref{hardness_lum}, unidentified sources
are shown by empty circles. The ROSAT count rates and upper limits are
adopted from the RASS Bright Source Catalog \citep{vab+99} and RASS Faint
Source Catalog except for V1062 Tau, where we used pointed ROSAT/PSPC
observations. The source XY Ari is not shown since its observed soft X-ray
flux is strongly diminished by absorption in an intervening molecular
cloud \citep{ldm01}. 
} 
\label{rxte_rosat_ratio}
\end{figure}

We expect our identified sample to be highly complete at $\lxh\gtrsim
10^{30}$~erg~s$^{-1}$ with respect to ABs and other types of coronal
stars for the following reasons. First, it is very unlikely that more
than $\sim$1--2 of the 13 unidentified XSS sources (see
Table~\ref{xss_noids}) for which there is no obvious bright
counterpart in the Rosat All-Sky Survey (RASS) are ABs or YSs, because
coronal X-ray sources are relatively soft. To illustrate this point we
plot in Fig.~\ref{rxte_rosat_ratio} the ratio of the ROSAT/PSPC count rate 
(0.1--2.4~keV) to the RXTE/PCA count rate (3--20~keV) as a function of
the latter for our identified and unidentified sources. For the 13 XSS
sources without a firm RASS counterpart an upper limit is
shown that was derived from the ROSAT/PSPC count rate of the
brightest RASS source within the XSS localization region (typically
0.5--1$^\circ$ in radius, R04). The unidentified XSS
sources without a bright RASS counterpart are apparently hard X-ray
sources compared to the identified ABs. It is important to note that the
presented XSS source fluxes are averages over multiple RXTE/PCA
observations separated by up to several years, hence it can be expected
that these fluxes are not strongly biased by individual X-ray
flares relative to the level of source persistent activity.

\begin{figure}
\centering
\includegraphics[width=\columnwidth]{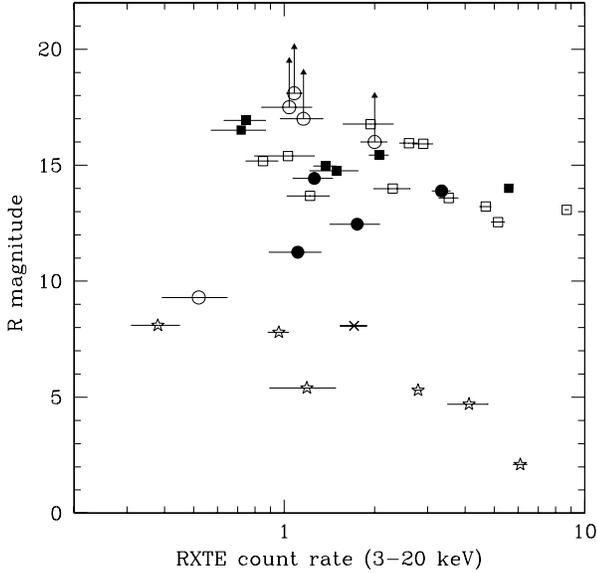}
\caption{$R$-band visual magnitude vs. RXTE/PCA count rate for the
identified XSS sources and 5 unidentified XSS sources with a firm RASS
counterpart. Identified sources of different classes are denoted by
the same symbols as in Fig.~\ref{hardness_lum}, unidentified sources
are shown by empty circles. $R$ magnitudes and lower limits are
adopted from the USNO-B1.0 Catalog. The XSS source XY Ari is not shown
since its optical spectrum is strongly reddened by absorption within a
molecular cloud \citep{ldm01}. 
} 
\label{rxte_optical}
\end{figure}

Secondly, for the 5 unidentified XSS sources reliably associated with
a RASS source (see Table~\ref{xss_noids}) we can search for a bright
star inside the ROSAT localization region (typically less than
30~arcsec in radius). A source with a 3--20~keV luminosity of
$10^{30}$--$10^{31.5}$~erg~s$^{-1}$ (the higher value is quite extreme
for coronal sources) would typically be detectable in the XSS out to
$\sim 20$--100~pc. Stars exhibiting such high levels 
of coronal activity are rapidly rotating (usually in short-period binaries)
main-sequence or evolved late-type stars, with $M_V\lesssim 6$ (see
e.g. \citealt{sdw96,m03}). Therefore, if any of the unidentified XSS
sources were a high luminosity coronal source, we would expect to 
find a star brighter than $V\sim 11$ in the ROSAT localization
region. Search of the Hipparcos and Tycho catalogs revealed only one
such bright star, a possible counterpart to XSS J17223$-$7301/1RXS
J171850.0$-$732527 (see Table~\ref{xss_noids}). Should this
association be confirmed, it will not significantly change our
estimate of the space density of ABs. Fig.~\ref{rxte_optical}
illustrates the above argument by showing the $R$-band visual
magnitudes (or lower limits) vs. the RXTE/PCA count rate for identifed
XSS sources and for the 5 unidentified XSS sources with a RASS
counterpart. One can see that the optical counterparts of the
unidentified XSS sources (except for XSS J17223$-$7301 mentioned
above) are much dimmer than expected for 
coronal sources.   

On the other hand, since CVs can be undetectable in the RASS due to
their hard spectra (see Fig.~\ref{rxte_rosat_ratio}) and can also be 
inconspicuous optically (see Fig.~\ref{rxte_optical}),  it is possible
that some of the unidentified XSS sources belong to this class. There
is an additional possibility to test the XLF obtained. Our identified
sample is highly complete in the northern hemisphere: there are 16
identified and 2 unidentified sources at $\delta>0$. This
contrasts with the southern hemisphere, where there are 13 (excluding
the high-luminosity XSS J17309$-$0552/RXS J173021.5$-$055933)
identified vs. 16 unidentified sources. It is therefore 
worth comparing XLFs determined from the northern and southern
subsamples. As shown in Fig.~\ref{lumfunc_rxte}, the resulting XLFs
agree with each other and with the all-sky XLF within the
uncertainties, although there is a hint that the southern sample of
CVs may be somewhat incomplete. 

We conclude that we may somewhat underestimate the combined XLF
of ABs and CVs at $\lxh\gtrsim 10^{31}$~erg~s$^{-1}$ since there may
remain several unidentified CVs in the XSS catalog. The associated
systematic uncertainty is unlikely to exceed 50\% though.

\section{Low luminosity range: Rosat All-Sky Survey}
\label{rass}

The weakest X-ray source (a candidate AB) in the XSS sample has a luminosity
$\lx\approx 10^{30}$~erg~s$^{-1}$ in the 2--10~keV band. To
extend our study to $\lx<10^{30}$~erg~s$^{-1}$ we need a large-area
survey that would be more sensitive than the XSS and highly complete
with respect to source identification. Since no such 
survey has been performed so far in the standard (2--10~keV) or
similar X-ray band, we should consider different options. Taking into account
the fact that coronal stars with $\lx<10^{30}$~erg~s$^{-1}$ are
characterized by soft spectra, such possibility is provided by the Rosat
All-Sky Survey. Furthemore, by using spectral data from different X-ray
missions it should be possible to convert the space density of ROSAT
sources detected in the 0.1--2.4~keV band to harder X-ray bands. We
follow this approach below.

Our analysis will be based on two published catalogs derived from the RASS:
the catalog of 100 most luminous X-ray stars within 50~pc of the Sun
(\citealt{m03}, hereafter M03) and the RASS catalog of the nearby
stars (\citealt{hss+99}, hereafter H99). The first catalog includes
all stars with 0.1--2.4~keV luminosity ($\lxs$) higher than $9.8\times
10^{29}$~erg~s$^{-1}$. The second catalog includes all objects from
the Third Catalog of Nearby Stars \citep{gj91} that were detected 
in the RASS. Both catalogs are well suited for our statistical
study since they are expected to be highly complete and since they
provide accurate parallax distances for the sources.


\begin{table*}
\begin{center}
\caption{Space densities of soft X-ray active stars derived from the RASS
\label{rass_info}
}
\smallskip

\begin{tabular}{crrcrcc}
\hline
\hline
\multicolumn{1}{c}{$\log\lxs$} &
\multicolumn{1}{c}{D} &
\multicolumn{2}{c}{All stars} &
\multicolumn{2}{c}{ABs} &
\multicolumn{1}{c}{Catalog}
\\   
\cline{3-4}
\cline{5-6}
\multicolumn{1}{c}{} &
\multicolumn{1}{c}{pc} &
\multicolumn{1}{c}{Number} &
\multicolumn{1}{c}{Density, pc$^{-3}$} &
\multicolumn{1}{c}{Number} &
\multicolumn{1}{c}{Density, pc$^{-3}$} &
\multicolumn{1}{c}{} 
\\
\hline
31.0--31.5 & 50 &   7 & $(1.3\pm 0.5)\times 10^{-5}$ &  6 & 
$(1.1\pm 0.5)\times 10^{-5}$ & M03 \\   
30.5--31.0 & 50 &  18 & $(3.4\pm 0.8)\times 10^{-5}$ & 12 &
$(2.3\pm 0.7)\times 10^{-5}$ & M03 \\   
30.0--30.5 & 50 &  73 & $(1.5\pm 0.2)\times 10^{-4}$ & 24 &
$(4.6\pm 0.9)\times 10^{-5}$ & M03 \\   
29.5--30.0 & 25 &  31 & $(4.7\pm 0.9)\times 10^{-4}$ &  9 &
$(1.4\pm 0.5)\times 10^{-4}$ & H99 \\ 
 ...       & 15 &     &                              &  3 &
$(2.1\pm 1.2)\times 10^{-4}$ & H99 \\ 
29.0--29.5 & 25 &  96 & $(1.5\pm 0.2)\times 10^{-3}$ & 11 &
$(1.7\pm 0.5)\times 10^{-4}$ & H99 \\ 
 ...       & 15 &     &                              &  7 &
$(5.0\pm 1.9)\times 10^{-4}$ & H99 \\
28.5--29.0 & 25 & 131 & $(2.0\pm 0.2)\times 10^{-3}$ &  5 &
$(8.0\pm 3.0)\times 10^{-5}$ & H99 \\ 
 ...       & 15 &     &                              &  3 &
$(2.1\pm 1.2)\times 10^{-4}$ & H99 \\ 
28.0--28.5 & 20 & 132 & $(3.9\pm 0.3)\times 10^{-3}$ &    &
                             & H99 \\ 
27.5--28.0 & 11 &  48 & $(8.6\pm 1.2)\times 10^{-3}$ &    &
                             & H99 \\ 
27.0--27.5 &  6 &  18 & $(2.0\pm 0.5)\times 10^{-2}$ &    &
                             & H99 \\ 
\hline             
\end{tabular}

\end{center}

\end{table*}


ABs of RS CVn, BY Dra, Algol, W Uma and other types (mostly of the
first two types) make up 43\% of the M03 sample. Another 42\% consist
mostly of pre-main-sequence and young main-sequence stars, while 15\%
of the stars are not classified. We may therefore determine the space
density of all sources and separately that of ABs.  Since the M03 is
volume limited, the source space density can be found as
\beq
\rho=\frac{N}{(4\pi/3) D^3},
\label{rho_volume}
\eeq
where $D=50$~pc for the M03 sample. We ignore here the small effect of
decreasing space density with height above the Galactic plane. The
resulting space densities for 3 luminosity intervals 
are given in Table~\ref{rass_info}. Note that although we ignored the
small number of unknown-type sources when estimating the space density
of ABs, this cannot significantly affect the result.

The H99 catalog is expected to be highly complete within 25~pc of the
Sun with respect to X-ray stars with $\lxs>1.5\times
10^{28}$~erg~s$^{-1}$. This follows from the fact that for 97\% of the
sky an exposure of 100~s or longer was achieved in the RASS
\citep{vab+99}, which for coronally active stars typically
corresponds to a 0.1--2.4~keV flux limit of $\sim 2\times
10^{-13}$~erg~cm$^{-2}$~s$^{-1}$ \citep{hss+99}. Given this flux limit
one can readily find a distance $D$ within which the H99 catalog should be
complete for a given limiting luminosity. One can then again apply
equation~(\ref{rho_volume}) to estimate the space density of X-ray
stars with luminosities exceeding this limit within distance $D$. 

To separate ABs from other sources we need information about 
source classes, which is not provided by H99. We hence
cross-correlated the H99 sample with published catalogs of
chromospherically active binaries \citep{shf+93,kbe+04}. A few
additional RS CVn and W Uma systems were 
found by cross-correlating the H99 catalog with the General Catalog of
Variable Stars \citep{sd+04}. Since it is possible that these catalogs
are not complete at low luminosities, we resticted our analysis to ABs
with $\lxs>10^{28.5}$~erg~s$^{-1}$. We additionally compared the space
density of ABs within 25~pc with that within
15~pc. Table~\ref{rass_info} provides space densities of X-ray stars
in a number of luminosity intervals, as derived from the H99
sample. 

\begin{figure}
\centering
\includegraphics[width=\columnwidth]{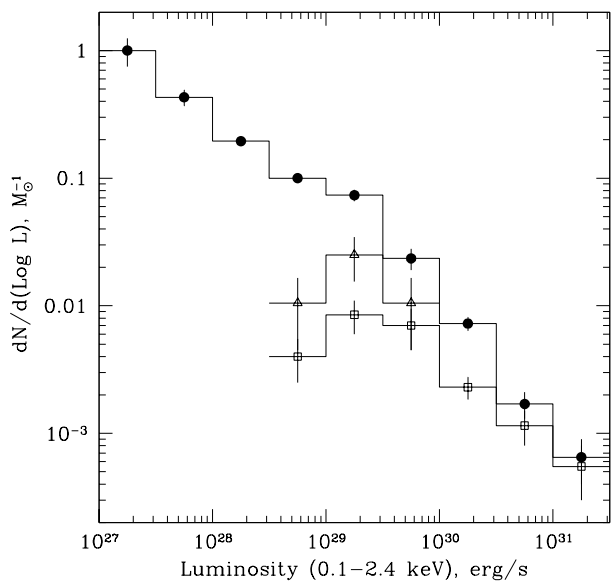}
\caption{Differential soft X-ray luminosity function of stars in the
solar neighborhood derived from the RASS. The filled circles show the
XLF of all stars, and the open squares show the XLF of ABs within
50~pc of the Sun. For three low-luminosity bins also the XLF of ABs
within 15~pc of the Sun is shown (triangles).  
} 
\label{lumfunc_rosat}
\end{figure}

Combining the results from the M03 and H99 samples we obtain
the differential soft X-ray luminosity function of nearby
low-luminosity sources shown in Fig.~\ref{lumfunc_rosat}. One can see
that between $\lxs\sim 10^{30.5}$ and $\sim 10^{31.5}$~erg~s$^{-1}$
ABs dominate the local X-ray source population. At lower luminosities, 
the fraction of YSs (and normal, Sun-like stars at the lower end of
the luminosity function) becomes progressively higher, although 
we point out that the somewhat higher space density of ABs found
within $D=15$~pc compared to $D=25$~pc may indicate that the catalogs of ABs 
start to be incomplete at $\lxs\lesssim 10^{29.5}$~erg~s$^{-1}$ within the
larger volume.

Our analysis implies that the local space density of ABs with 
$\lxs>10^{28.5}$~erg~s$^{-1}$ is $(4.7\pm 0.8)\times
10^{-4}$~pc$^{-3}$ (using $D=25$~pc for the H99 sample), in
satisfactory agreement with the estimate based on the  
Einstein Extended Medium Sensitivity Survey of $(2.9\pm 0.6)\times
10^{-4}$~pc$^{-3}$ \citep{fgm89,os92}. It also follows from the above 
analysis that stars with $\lxs<10^{29}$~erg~s$^{-1}$ produce less than
20\% of the total local soft X-ray emissivity (see
Table~\ref{emissivities}). We point out that the soft X-ray luminosity
function shown in Fig.~\ref{lumfunc_rosat} 
extends from the most luminous coronal stars ($\lxs\sim
10^{31.5}$~erg~s$^{-1}$) down to Sun-like stars ($\lxs\sim
10^{27}$~erg~s$^{-1}$).    

\subsection{Conversion from the ROSAT energy band to the standard X-ray band}

Conversion of the soft X-ray luminosity function obtained above to the
2--10~keV energy band requires knowledge of the source spectra. Since
the majority of RASS sources used in our analysis have not been
observed in the standard X-ray band, we are bound to rely on a
representative set of sources for which broad-band spectra are
available. To this end we selected from public archives X-ray
observations, of sufficiently good quality for spectral analysis, for 22
sources from the M03 sample and 25 sources from the H99 sample. All
observations were performed by ASCA, except for the star GJ 1245
observed by Chandra. ASCA and Chandra data were then processed by
standard tasks of HEASOFT and CIAO packages according to recipes of
the Guest Observer Facilities
(http://legacy.gsfc.nasa.gov/docs/asca/ascagof.html and
http://cxc.harvard.edu/ciao/). 

In the 0.5--10~keV band the (moderate resolution) spectra of
all selected sources are well fit by a broken power law with the
break energy and lower-energy photon index fixed at 0.8~keV and 1.5,
respectively. The high-energy photon index was a free parameter in our
analysis and we found for it best-fit values in the range from $\sim
3$ to $\sim 5$ for different sources. This simple empirical model
mimics reasonably well the actual multi-temperature emission spectrum
(e.g. \citealt{scs+90,dls+93,g04}) dominated by strong blended line 
emission below $\sim 0.8$~keV. From the best-fit model we can find for
each source the ratio of its luminosity in the 2--10~keV band to 
that in the 0.5--2~keV band.  

\begin{figure}
\centering
\includegraphics[width=\columnwidth]{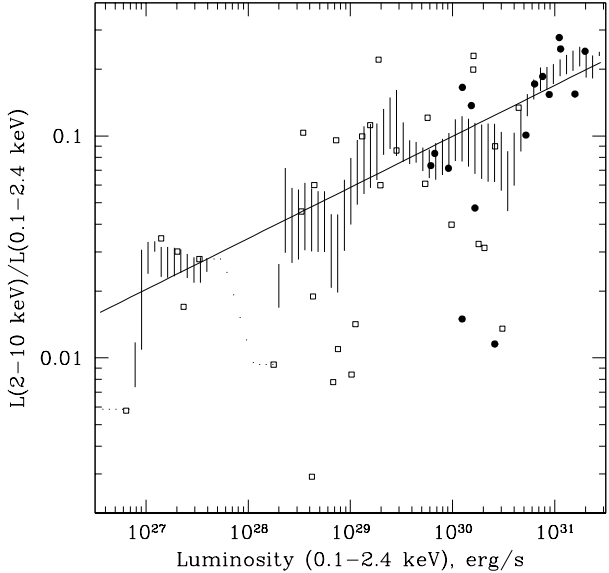}
\caption{Ratio of luminosities in the 2--10~keV and 0.1--2.4~keV
energy bands as a function of soft X-ray luminosity for RASS sources,
estimated from ASCA or Chandra spectra. Solid symbols denote ABs and 
open symbols other coronal stars. Also shown (shaded region) is the 1$\sigma$
confidence region (determined by the scatter of individual
measurements around the mean value) for a sliding-window average
($\Delta\log\lxs=0.5$) of the presented data points, and
the power-law approximation (solid line) to this average given by equation
(\ref{rosat_2_10}).   
} 
\label{ratio_rosat}
\end{figure}

We then additionally convert $L$~(0.5--2~keV) to $\lxs$, luminosity
in the ROSAT (0.1--2.4~keV) band, assuming $L$(0.5--2~keV)/$\lxs\sim
0.7$, a ratio typical for coronal stars observed by
Einstein and ROSAT \citep{fmm+95}. The resulting values of $\lx/\lxs$ are
plotted as a function of $\lxs$ in Fig.~\ref{ratio_rosat} for our
spectral sample of sources. In most cases the $\lxs$ value determined
from ASCA or Chandra  observations differs by less than a factor of 2
from the soft X-ray luminosity directly measured by ROSAT at a different epoch.
 
Fig.~\ref{ratio_rosat} confirms the well-known trend of hardening of
stellar coronal X-ray emission with increasing luminosity
(e.g. \citealt{scs+90,g04}). It also demonstrates that ABs do
not distinguish themselves spectrally from YSs with similar
luminosities. Averaging the ratios $\lx/\lxs$ for individual
sources in a sliding window of width $\Delta\log\lxs=0.5$ leads to the
result shown by the shaded region in Fig.~\ref{ratio_rosat}, which
reflects the uncertainty in $\langle\lx/\lxs\rangle$ due to the scatter
of individual $\lx/\lxs$ values around this mean value. The
sliding-window average can be approximated by the power law
\beq 
\langle\frac{\lx}{\lxs}\rangle=0.045\left(\frac{\lxs}{10^{28.5}}\right)^{0.23},
\label{rosat_2_10}
\eeq
shown by the solid line in Fig.~\ref{ratio_rosat}.

The width of the shaded region in Fig.~\ref{ratio_rosat} indicates
that the hardness-luminosity trend described by equation
(\ref{rosat_2_10}), which is based on a fairly small sample of sources, should
enable $\sim$50\% accuracy of conversion of our soft X-ray luminosity
function (Fig.~\ref{lumfunc_rosat}), derived from a much larger sample
of RASS sources, to the 2--10~keV energy band for  $\lxs\gtrsim
10^{28.5}$~erg~s$^{-1}$. 

We will also need below a similarly determined approximate
trend for the harder energy band 3--20~keV:
\beq
\langle\frac{\lxh}{\lxs}\rangle=0.02\left(\frac{\lxs}{10^{28.5}}\right)^{0.31}.
\label{rosat_3_20}
\eeq

\section{Combined X-ray luminosty function}
\label{broad}

\begin{figure}
\centering
\includegraphics[width=\columnwidth]{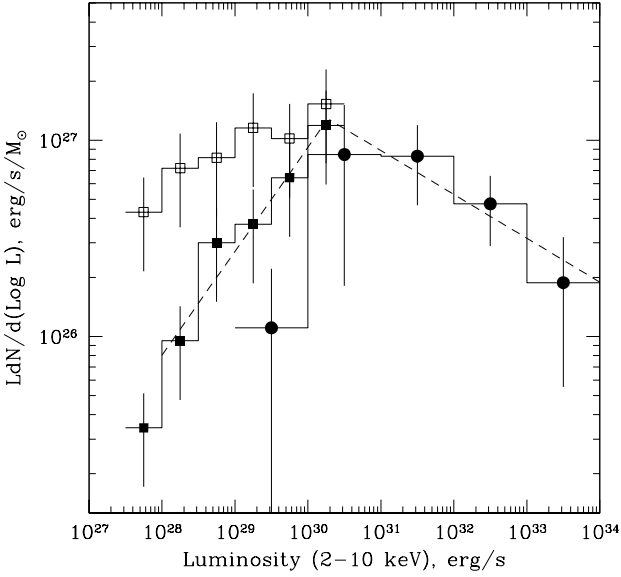}
\caption{Differential luminosity distribution of 2--10~keV emissivity
per unit stellar mass of coronally active stars and CVs. The XLF derived
from the XSS is shown by filled circles, and the XLF derived from the
RASS is shown by open squares for all stars and by filled
squares for ABs only. The errors shown for the RASS data points take
into account an assumed 50\% uncertainty of conversion from the original
0.1--2.4~keV band in addition to statistical errors. The dashed line
shows the broken power-law fit to the combined XLF of ABs and CVs,
given by equation~(\ref{xlf210_fit}). 
}
\label{lumfunc_en_2_10}
\end{figure}

\begin{figure}
\centering
\includegraphics[width=\columnwidth]{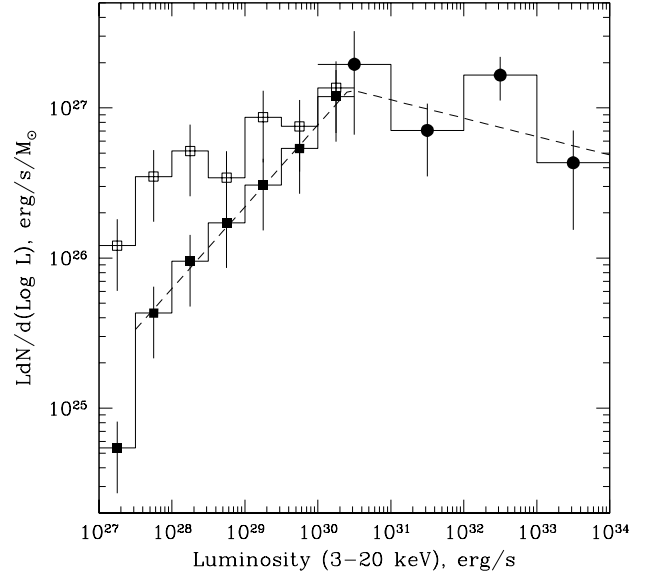}
\caption{Same as Fig.~\ref{lumfunc_en_2_10}, but for the 3--20~keV
band. The dashed line shows the best-fitting model given by
equation~(\ref{xlf320_fit}). 
}
\label{lumfunc_en_3_20}
\end{figure}

We now proceed to convert to a common energy band the X-ray
(3--20~keV) and soft X-ray (0.1--2.4~keV) luminosity functions derived
from the XSS and RASS in Section~\ref{xss} and Section~\ref{rass},
respectively. We first consider the 2--10~keV band. For the XSS sample
we can readily recompute the XLF using the 2--10~keV source
luminosities given in Table~\ref{xss_sources}. We apply a more
approximate procedure to the RASS sample, namely convert the measured soft
X-ray luminosities to the 2--10~keV range using the approximate
hardness-luminosity trend given by equation (\ref{rosat_2_10}) and
then recompute the XLF. The two recomputed XLFs make up a broad
range XLF (from $10^{27.5}$ to $10^{34}$~erg~s$^{-1}$) that is shown in
Fig.~\ref{lumfunc_en_2_10}. We can similarly construct an XLF in the
3--20~keV band (Fig.~\ref{lumfunc_en_3_20}); in this case only the
RASS XLF needed to be recomputed using
equation~(\ref{rosat_3_20}). The 3--20~keV XLF is used by \cite{r+05}
to assess the contribution of point sources to the Galactic  
ridge X-ray emission measured by RXTE in the same energy band.
 
The XLFs shown in Fig.~\ref{lumfunc_en_2_10} and
Fig.~\ref{lumfunc_en_3_20} were multipled by luminosity to expose the 
contribution of different luminosity intervals to the total X-ray
emissivity per unit stellar mass. In the low-luminosity range covered
by RASS data ($\lx<10^{30.5}$~erg~s$^{-1}$), both the total XLF
including YSs and separately that of ABs are
shown. To roughly allow for the uncertainty of conversion from the
original soft X-ray band to the 2--10~keV and 3--20~keV bands we ascribed 50\% 
errors to the RASS data points in addition to statistical uncertainties. 

The medium-luminosity XLF derived from the XSS and the low-luminosity
XLF derived from the RASS partially overlap near $10^{30}$~erg~s$^{-1}$, in a
region occupied predominantly by ABs, and do not contradict each
other. For the subsequent analysis we will adopt the XSS estimates of
differential source space densities in the $(10^{30},10^{34})$ 
luminosity range and the RASS estimates of space densities of
lower-luminosity sources.  

The combined 2--10~keV XLF of ABs and CVs can be 
approximated in the range $10^{28}$--$10^{34}$~erg~s$^{-1}$ by a
broken power law:   
\beqa
\frac{dN}{d\log\lx} &=&
K\left\{
\begin{array}{ll}
(L_{\rm b}/\lx)^{\alpha_1}, & \lx<L_{\rm b},\\
(L_{\rm b}/\lx)^{\alpha_2}, & \lx>L_{\rm b},\\
\end{array}
\right.
\label{xlf210_fit}
\eeqa
where $K\approx 6.8\times 10^{-4}$~$\ms^{-1}$, $L_{\rm
b}\approx 1.9\times 10^{30}$~erg~s$^{-1}$, $\alpha_1\approx 0.47$ and
$\alpha_2\approx 1.22$. We note that incompleteness may significantly
affect the XLF of ABs below $\lx\sim 10^{28}$~erg~s$^{-1}$
(corresponding to $\lxs\sim 10^{29.5}$~erg~s$^{-1}$, see
Section~\ref{rass}) and also somewhat the XLF of CVs above $\lx\sim
10^{31}$~erg~s$^{-1}$ (see Section~\ref{xss}). Similarly the 3--20~keV
XLF of ABs and CVs can be fitted in the range
$10^{27.5}$--$10^{34}$~erg~s$^{-1}$ by 
\beqa
\frac{dN}{d\log\lxh} &=&
K\left\{
\begin{array}{ll}
(L_{\rm b}/\lxh)^{\alpha_1}, & \lxh<L_{\rm b},\\
(L_{\rm b}/\lxh)^{\alpha_2}, & \lxh>L_{\rm b},\\
\end{array}
\right.
\label{xlf320_fit}
\eeqa
with $K\approx 4.9\times 10^{-4}$~$\ms^{-1}$, $L_{\rm
b}\approx 2.7\times 10^{30}$~erg~s$^{-1}$, $\alpha_1\approx 0.45$ and
$\alpha_2\approx 1.12$. These analytical fits (multiplied
by luminosity) are shown by dashed lines in Fig.~\ref{lumfunc_en_2_10}
and Fig.~\ref{lumfunc_en_3_20}.  
 
\subsection{Cumulative emissivities}
\label{cumemiss}

\begin{figure}
\centering
\includegraphics[width=\columnwidth]{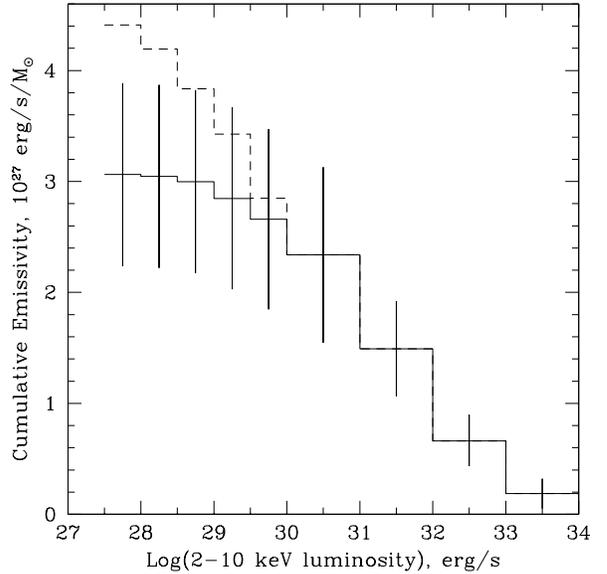}
\caption{Cumulative 2--10~keV emissivity of ABs and CVs as a function of
luminosity (solid histogram and error bars) and cumulative emissivity
of all coronal stars and CVs (dashed histogram). The error bars
translate from those for the XLF shown in Fig.~\ref{lumfunc_en_2_10}). 
}
\label{cumemiss_2_10}
\end{figure}

\begin{figure}
\centering
\includegraphics[width=\columnwidth]{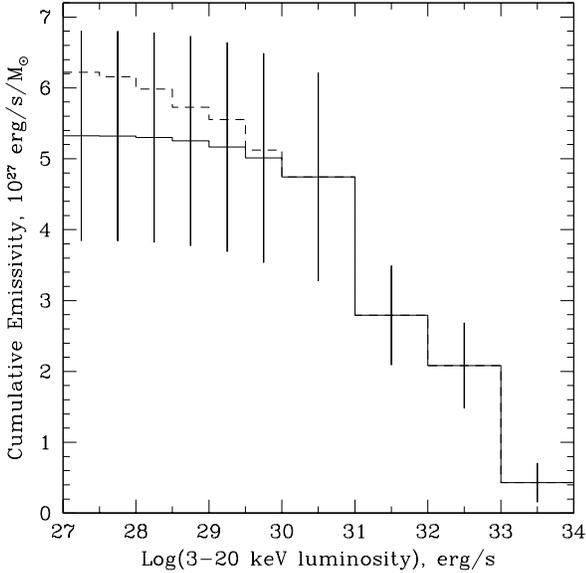}
\caption{Same as Fig.~\ref{cumemiss_2_10}, but for the 3--20~keV band.
}
\label{cumemiss_3_20}
\end{figure}

Using the differential XLFs obtained above we can assess the cumulative
emissivity of local X-ray sources with luminosities below  
$10^{34}$~erg~s$^{-1}$. We present in Fig.~\ref{cumemiss_2_10} and
Fig.~\ref{cumemiss_3_20} the corresponding plots for the
2--10~keV and 3--20~keV bands. Table~\ref{emissivities} summarizes our
estimates of the cumulative local emissivities (per unit stellar mass)
of ABs, CVs and YSs in the energy bands 0.1--2.4~keV, 2--10~keV and
3--20~keV, complemented by information about  LMXBs (see
Section~\ref{lmxb} below).

Approximately 80\% of the total X-ray (2--10~keV) luminosity of ABs and CVs is
produced by sources with $\lx>10^{30}$~erg~s$^{-1}$. In the
solar neighborhood an additional significant contribution comes from
YSs with $\lx\lesssim 10^{30}$~erg~$^{-1}$, which make up $\sim 30$\%
of the total luminosity at 2--10~keV. The fractional contribution of
YSs decreases when going to a harder X-ray band because of their
relatively soft spectra (compare Fig.~\ref{cumemiss_2_10} and
Fig.~\ref{cumemiss_3_20}). We note that the estimated (by integrating
the soft X-ray luminosity function shown in Fig.~\ref{lumfunc_rosat} up to
$\lxs=10^{32}$~erg~s$^{-1}$) high soft X-ray emissivities of ABs and
YSs compared to the harder X-ray bands reflect the fact that stellar
coronae are much more efficient sources of soft X-rays than hard
X-rays. We also point out that the lower luminosity end of the
distributions shown in Fig.~\ref{cumemiss_2_10} and
Fig.~\ref{cumemiss_3_20} corresponds to $\lxs\sim
10^{29}$~erg~s$^{-1}$ and the contribution of less luminous X-ray 
stars (including normal stars like the Sun) to the total X-ray
emissivity above 2~keV is expected to be negligible since they contribute less
than 20\% to the soft X-ray emissivity and are softer than the more
luminous sources (see Section~\ref{rass}). 


\begin{table*}
\begin{center}
\caption{Number densities and emissivities of diffirent classes
of sources
\label{emissivities}
}
\smallskip

\begin{tabular}{ccccc}
\hline
\hline
\multicolumn{1}{c}{Class} & 
\multicolumn{1}{c}{Space density above given luminosity} &
\multicolumn{3}{c}{Total emissivity ($10^{27}$~erg~s$^{-1}$~$\ms^{-1}$)}
\\
\cline{3-5}
\multicolumn{1}{c}{} &
\multicolumn{1}{c}{ ($\ms^{-1}$)} &
\multicolumn{1}{c}{0.1--2.4~keV} &
\multicolumn{1}{c}{2--10~keV} &
\multicolumn{1}{c}{3--20~keV}
\\   
\hline
ABs         & $\sim 1.2\times 10^{-2}$ ($\lx>10^{27.5}$~erg~s$^{-1}$) &
 $(14\pm 4)$ & $2.0\pm 0.8$ & $2.9\pm 1.3$\\  
CVs         & $(1.2\pm 0.3)\times 10^{-5}$ ($\lx>10^{31}$~erg~s$^{-1}$)
& $\lesssim${\rm a few} & $1.1\pm 0.3$ & $2.4\pm 0.6$\\
ABs+CVs     & & $\sim 15$        & $3.1\pm 0.8$  & $5.3\pm 1.5$\\
 YSs        & & $24\pm 3$        & $1.5\pm 0.4$  & $1.0\pm 0.2$\\
ABs+CVs+YSs & & $\sim 40$        & $4.5\pm 0.9$  & $6.2\pm 1.5$\\
LMXBs   & $\sim 3\times 10^{-9}$  ($\lx>10^{36}$~erg~s$^{-1}$) &
  $\sim 40$$^{\rm a}$ & $\sim 90$$^{\rm b}$    &   \\
\hline

\end{tabular}

\end{center}

$^{\rm a}$ Emissivity in the 0.5--2~keV band extrapolated from
2--10~keV assuming a power-law spectrum of photon index $\Gamma=1.56$,
as typical of LMXBs with
$10^{36}$~erg~s$^{-1}<\lx<10^{39}$~erg~s$^{-1}$ \citep{iab03}.

$^{\rm b}$ The quoted 2--10~keV emissivity for LMXBs represents an
average over nearby galaxies \citep{gilfanov04}. 

\end{table*}


Our preceeding analysis does not permit to estimate the soft X-ray
emissivity of CVs. The XSS sample is not suitable for this purpose
because the high-energy component (optically thin thermal emission
with $kT\lesssim 30$~keV) of CV spectra observed by RXTE or a
similar X-ray instrument is often intrinsically absorbed below
several keV (e.g. \citealt{crw98,srr05}), while another, much softer
component (black-body emission with $kT\sim 30$~eV) appears in the
ROSAT energy range, with the relative amplitudes of the two components
varying greatly from source to source (\citealt{c90}, see also
Fig.~\ref{rxte_rosat_ratio}). Therefore, to obtain a reliable estimate of
the CV soft X-ray emissivity one has to use a flux limited and
optically identified soft X-ray survey such as the Rosat Bright Survey
(RBS, \citealt{shl+00}).   

The RBS was already used by \cite{sbb+02} to estimate the space
density of non-magnetic CVs. Using the same sample of 15 non-magnetic CVs
with measured distances (Table~4 in \citealt{sbb+02}, which provides
source luminosities and $\vmm$ values) we can readily estimate the
soft X-ray cumulative emissivity (per unit stellar mass) of
non-magnetic CVs: $\sim 7\times 10^{26}$~erg~s$^{-1}$.

Unfortunately, as noted by \cite{sbb+02}, the RBS sample of magnetic
CVs substantially  suffers from incomplete distance information, which
currently makes diffucult its use for statistical studies. Using the published
estimate of the space density of magnetic CVs of $\sim 3\times
10^{-7}$~pc$^{-3}$ \citep{p84,w95}, which may be affected by different
biases but nevertheless agrees with our XSS based estimate of space
density of magnetic CVs with $\lx>10^{31}$~erg~s$^{-1}$ of $(4.8\pm
1.6)\times 10^{-7}$~pc$^{-3}$, and assuming $\lxs\sim 5\times
10^{31}$~erg~s$^{-1}$ for the average source luminosity
(e.g. \citealt{bsm99}), we can estimate the soft X-ray emissivity of
magnetic CVs at $\sim 4\times
10^{26}$~erg~s$^{-1}$~$\ms^{-1}$. Considering that this estimate can
be inaccurate by a factor of a few, we infer that the combined soft X-ray  
emissivity of non-magnetic and magnetic CVs is likely less than a few
$10^{27}$~erg~s$^{-1}$~$\ms^{-1}$. This implies that the total local
soft X-ray emissivity is strongly dominated by ABs and YSs (see
Table~\ref{emissivities}). 

\subsection{Addition of LMXBs}
\label{lmxb}

\begin{figure}
\centering
\includegraphics[width=\columnwidth]{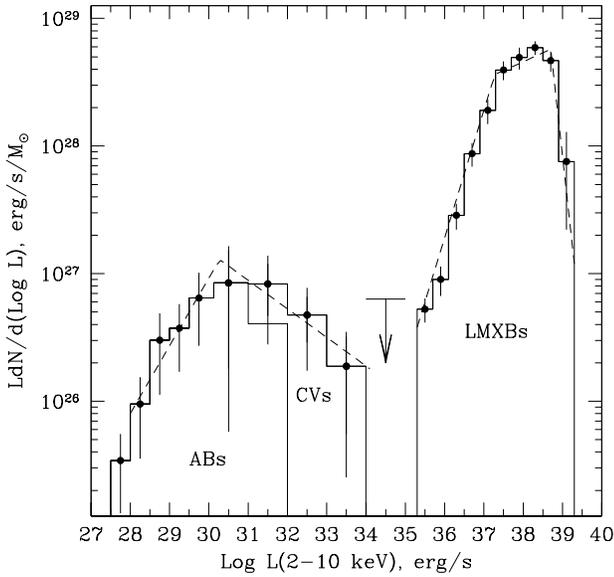}
\caption{Differential luminosity distribution of 2--10~keV emissivity of
ABs, CVs and LMXBs. Also the contributions of these classes of sources
are indicated and analytical approximations given by
equation~(\ref{xlf210_fit}) and \cite{gilfanov04} are presented 
(dashed lines). 
}
\label{lumfunc_broad}
\end{figure}

The XLF of Galactic LMXBs in the energy band 2--10~keV was constructed by
\cite{ggs02}. \cite{gilfanov04} subsequently demonstrated that the LMXB XLFs 
for 11 nearby galaxies and the Milky Way have a universal shape and
normalizations proportional to the stellar masses. We can now attach to the XLF
of high luminosity LMXBs ($\sim 10^{35}$--$\sim
10^{39}$~erg~s$^{-1}$) averaged over nearby galaxies the XLF of ABs
and CVs constructed here. The combined XLF (per unit
stellar mass) multiplied by luminosity is shown in
Fig.~\ref{lumfunc_broad}. 
 
The only remaining poorly studied luminosity interval is
$10^{34}$--$\sim 10^{35}$~erg~s$^{-1}$, but it is possible to place an
upper limit on the space density of  objects with
such luminosities based on the ASCA Galactic Plane Survey
\citep{smk+01}. This survey covered $\approx 40$~sq. deq within the
central region of the Galactic plane ($|l|< 45^\circ$ and $|b|<
0.4^\circ$) with the flux limit in the 2--10~keV
energy band varying between $\sim 10^{-12.5}$ and
$\sim 10^{-12}$~erg~cm$^{-2}$~s$^{-1}$. 

It follows from the number-flux distribution obtained by
\cite{smk+01} that there are on average $\sim$4~deg$^{-2}$ Galactic
sources with flux higher than $10^{-12.5}$~erg~cm$^{-2}$~s$^{-1}$
within the region $|l|< 45^\circ$, $|b|< 0.4^\circ$. This implies that
the total number of such sources in this region is $\sim
290$. Since the vast majority of weak sources detected in the ASCA 
survey are unidentified, we can conservatively assume that all
detected sources brighter than $10^{-12.5}$~erg~cm$^{-2}$~s$^{-1}$
have luminosities exceeding $10^{34}$~erg~s$^{-1}$. At the flux limit of
the survey, a source with $\lx>10^{34}$~erg~s$^{-1}$ is detectable 
out to a distance $>16$~kpc, i.e. almost to the outer boundary of the
Galactic disk. Using the model of stellar mass distribution in the
Galactic disk   
\beq
\rho\propto\exp\left[-\left(\frac{R_{\rm m}}{R}\right)^3-\frac{R}{R_{\rm
scale}}-\frac{z}{z_{\rm scale}}\right],
\label{disk}
\eeq
we find that $\approx 30$\% of the total mass of the disk is
contained within $|l|< 45^\circ$, $|b|< 0.4^\circ$. Here we have
assumed $R_{\rm m}=3$~kpc, $R_{\rm scale}=3$~kpc, $R_{\rm
max}=10$~kpc and $z_{\rm scale}=150$~pc
\citep{binney97,freudenreich98}, although the result is almost
insensitive to the parameter values except for the scale height 
$z_{\rm scale}$. Taking additionally into account that $\sim 30$\% of
the Milky Way stellar mass is contained in the bulge and halo
\citep{bahcall80,freudenreich98}, which are virtually
not covered by the ASCA Galactic Plane Survey, we may 
conservatively estimate that there are less than $290/0.3/(1-0.3)\sim
1400$ sources with $10^{34}$~erg~s$^{-1}<\lx<10^{35}$~erg~s$^{-1}$ in
the Galaxy. Adopting the value $7\times 10^{10}\ms$ for the mass of the
Galaxy in stars (derived from the K-band luminosity measured by COBE,
\citealt{msr+96,gilfanov04}), we finally obtain the upper limit shown in
Fig.~\ref{lumfunc_broad}.   

It can be seen from Fig.~\ref{lumfunc_broad} that the differential
luminosity distribution of X-ray emissivity of Galactic
low-mass close binaries has two maxima. The primary peak at $\lx\sim
10^{38}$~erg~s$^{-1}$ is due to neutron-star LMXBs 
accreting at near the Eddington limit. The secondary peak, at $\sim
10^{29}$--$10^{33}$~erg~s$^{-1}$, is formed jointly by ABs and
CVs. The XLF can be approximated by equation (\ref{xlf210_fit})
in the range $10^{28}$--$10^{34}$~erg~s$^{-1}$ and by the LMXB
template given in \cite{gilfanov04} [eqs. (8),(9) and Table~(3)]
in the range $10^{35}$--$10^{39}$~erg~s$^{-1}$. Both
analytical fits are shown by dashed lines in Fig.~\ref{lumfunc_broad}.

\begin{figure}
\centering
\includegraphics[width=\columnwidth]{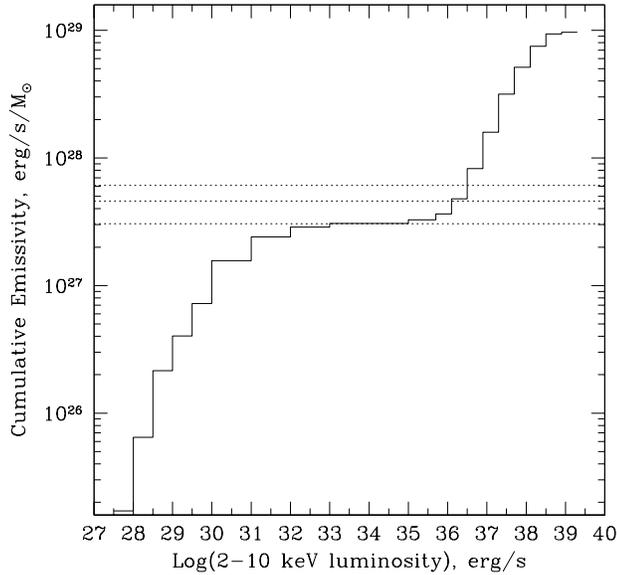}
\caption{Cumulative 2--10~keV emissivity (computed from low luminosities
upward) of ABs, CVs and LMXBS as a function of luminosity. The dashed
lines show the levels of 100\%, 150\% and 200\% of the total emissivity of
ABs and CVs.
}
\label{cumemiss_broad_2_10_rev}
\end{figure}

In Fig.~\ref{cumemiss_broad_2_10_rev} we show the cumulative 2--10~keV
emissivity of ABs, CVs and LMXBs as a function of luminosity. LMXBs
provide by far the dominant contribution ($\sim
10^{29}$~erg~s$^{-1}$~$\ms^{-1}$) to the total emissivity, whereas ABs
and CVs together contribute $\sim 3$\%. Fig.~\ref{cumemiss_broad_2_10_rev} also
demonstrates the effect of cutting out the bright end of the combined
XLF: the cumulative emissivity of LMXBs with $\lx<10^{36}$
($\lx<10^{36.5}$)~erg~s$^{-1}$ is $\sim 50$\% ($\sim 100$\%) of the
total emissivity of ABs and CVs.   

\begin{figure}
\centering
\includegraphics[width=\columnwidth]{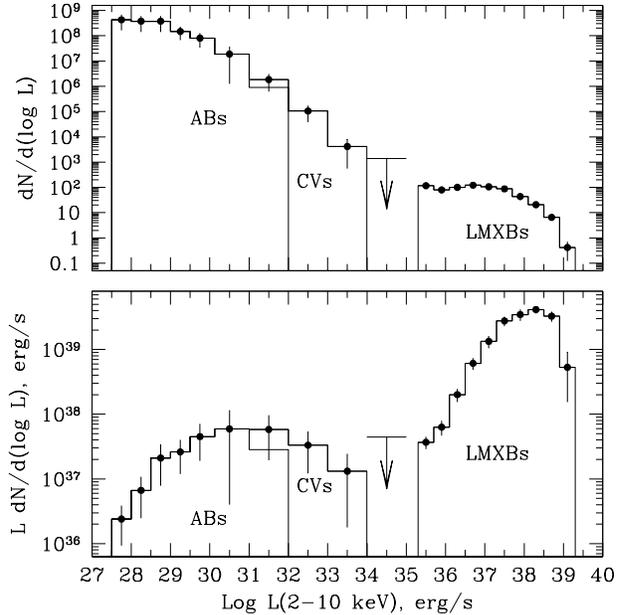}
\caption{{\sl Upper panel:} XLF of Galactic low-mass close binaries. The
Galaxy stellar mass is assumed to be $7\times 10^{10}$~$\ms$ and the
LMXB part of the XLF is averaged over nearby galaxies \citep{gilfanov04}. {\sl
Lower panel:} Luminosity distribution of X-ray energy output of Galactic
low-mass X-ray binaries.
}
\label{lumfunc_final}
\end{figure}

Finally Fig.~\ref{lumfunc_final} shows the predicted XLF and the
luminosity distribution of X-ray energy output of ABs, CVs
and LMXBs for the entire Galaxy. The predicted contribution
from ABs and CVs to the 2--10 keV luminosity of the Milky Way is $\sim
2\times 10^{38}$~erg~s$^{-1}$, which agrees within the measurement
uncertainties with the total X-ray luminosity of the Galactic ridge
X-ray emission (see a detailed discussion in \citealt{r+05}). 

\section{Conclusions}
\label{conclusions}

In this paper we have constructed the X-ray (above 2 keV)
luminosity function of coronally active binaries, CVs and LMXBs,
covering $\sim 12$ orders of magnitude in luminosity.

We find that the differential luminosity distribution of X-ray emissivity (per
unit stellar mass) of low-mass close binaries has a broad secondary
peak at $\lx\sim 10^{29}$--$10^{33}$~erg~s$^{-1}$ composed of ABs and
CVs, in addition to the previously well studied primary maximum at
$\sim 10^{38}$~erg~s$^{-1}$ made up by neutron-star LMXBs accreting at
near the Eddington limit. The combined emissivity of ABs and CVs in
the 2--10~keV band is $(3.1\pm 0.8)\times
10^{27}$~erg~s$^{-1}$~$\ms^{-1}$, or $\sim 3$\% of the emissivity of
LMXBs (averaged over nearby galaxies). About 65\% of this total emissivity
is due to ABs. The estimated combined contribution 
of ABs and CVs to the 2--10~keV luminosity of the Milky Way is
$\sim 2\times 10^{38}$~erg~s$^{-1}$. 

Young coronal stars with luminosities $\lx\lesssim
10^{30}$~erg~s$^{-1}$ provide an additional significant contribution
of $(1.5\pm 0.4)\times 10^{27}$~erg~s$^{-1}\ms^{-1}$ to 
the cumulative 2--10~keV emissivity in the solar neighborhood 
(within $\sim 50$~pc). However, the fractional contribution of YSs to
the X-ray emissivity is expected to vary substantially across the
Galaxy, reflecting local star formation history. In contrast, the
cumulative X-ray emission of ABs and CVs is expected to approximately
follow the distribution of stellar mass, as is known to be the case for LMXBs. 

The results of this work find immediate application to the problem of
the origin of Galactic ridge X-ray emission. \cite{r+05} use the XLF
constructed here in combination with the X-ray surface brightness
distribution of the ridge emission, which is shown to follow the
stellar mass, to demonstrate that ABs and CVs (with a possible
contribution from YSs) likely produce the bulk of the
ridge emission.

The results of this work also indicate that in order to assess contribution of
low-luminosity point X-ray sources (ABs and CVs) to the apparently
diffuse X-ray emission of gas poor elliptical galaxies, it is
necessary to resolve out the brightest LMXBs with $\lx\gtrsim 
10^{36}$~erg~s$^{-1}$ (see Fig.~\ref{cumemiss_broad_2_10_rev}). This
can already be achieved with Chandra for nearby galaxies. It
should be taken into account however that in elliptical galaxies a
significant fraction of low-mass close binaries reside
in globular clusters where their numbers are expected to be affected
by dynamical processes in combination with aging (e.g. \citealt{wl05}). It
will be important to compare in future work the XLF derived here for the 
solar neighborhood with that determined for Galactic globular clusters
from deep Chandra observations (e.g. \citealt{hge+05}). 

\smallskip
\noindent {\sl Acknowledgments} This research has made use of the
SIMBAD database (operated at CDS, Strasbourg) and the High Energy
Astrophysics Science Archive Research Center Online Service provided
by the NASA/Goddard Space Flight Center.  



\begin{thebibliography}{}

\bibitem[\protect\citeauthoryear{Araujo-Betancor et
al.}{2005}]{agl+05} Araujo-Betancor, S., G\"{a}nsicke, B.T., Long,
K.S., Beuermann, K., de Martino, D., Sion, E.M., \& Szkody, P. 2005,
ApJ, 622, 589

\bibitem[\protect\citeauthoryear{Bahcall \& Soneira}{1980}]{bahcall80}
Bahcall J.~N., Soneira R.~M., 1980, ApJS, 44, 73

\bibitem[\protect\citeauthoryear{Barrett et al.}{1999}]{bsm99}
Barrett, P., Singh, K.P., \& Mitchell, S. 1999, Annapolis Workshop on
Magnetic Cataclysmic Variables, ASP Conference Series, Vol. 157,
eds. Coel Hellier \& Koji Mukai, p. 180  

\bibitem[\protect\citeauthoryear{Baskill et al.}{2005}]{bwo05}
Baskill, D.S., Wheatley, P.J., \& Osborne, J.P. 2005, MNRAS, 357, 626

\bibitem[\protect\citeauthoryear{Beuermann et al.}{2004}]{bhm+04}
Beuermann, K., Harrison, Th.E., McArthur, B.E., Benedict, G.F., \&
G\"{a}nsicke, B.T. 2004, A\&A, 419, 291
	
\bibitem[\protect\citeauthoryear{Binney, Gerhard \&
Spergel}{1997}]{binney97} Binney J., Gerhard O., \& Spergel D. 1997,
MNRAS, 288, 365 

\bibitem[\protect\citeauthoryear{Canizares et al.}{1987}]{cft87}
Canizares, C.R., Fabbiano, G., \& Trinchieri, G. 1987, ApJ, 312, 503

\bibitem[\protect\citeauthoryear{Cropper}{1990}]{c90} Cropper,
M. 1990, SSRv, 54, 195

\bibitem[\protect\citeauthoryear{Cropper et al.}{1998}]{crw98}
Cropper, M., Ramsay, G., \& Wu, K. 1998, MNRAS, 293, 222

\bibitem[\protect\citeauthoryear{Dempsey et al.}{1993}]{dls+93}
Dempsey, R.C., Linsky, J.L., Schmitt, J.H.M.M., \& Fleming, T.A. 1993,
ApJ, 413, 333

\bibitem[\protect\citeauthoryear{Ebisawa et al.}{2005}]{etp+05}
Ebisawa, K., et al. 2005, ApJ (in press)

\bibitem[\protect\citeauthoryear{Eisenbart et al.}{2002}]{ebr+02}
Eisenbart, S., Beuermann, K., Reinsch, K., \& G\"{a}nsicke, B.T. 2002,
A\&A, 382, 984
		
\bibitem[\protect\citeauthoryear{Fleming et al.}{1989}]{fgm89} Fleming,
T.A., Gioia, I.M., \& Maccacaro, T. 1989, AJ, 98, 692

\bibitem[\protect\citeauthoryear{Fleming et al.}{1995}]{fmm+95}
Fleming, T.A., Molendi, S., Maccacaro, T., \& Wolter, A. 1995, ApJS,
99, 701

\bibitem[\protect\citeauthoryear{Freudenreich}{1998}]{freudenreich98}
Freudenreich H.T. 1998, ApJ, 492, 495

\bibitem[\protect\citeauthoryear{G\"{a}nsicke et al.}{2005}]{gme+05}
G\"{a}nsicke, B.T., Marsh, T.R., Edge, A., Rodr\'{i}guez-Gil, P.,
Steeghs, D., Araujo-Betancor, S., et al. 2005, MNRAS, 361, 141

\bibitem[\protect\citeauthoryear{Gilfanov}{2004}]{gilfanov04} Gilfanov,
M. 2004, MNRAS, 349, 146

\bibitem[\protect\citeauthoryear{Gliese \& Jahrei\ss}{1991}]{gj91}
Gliese, W., \& Jahrei\ss, H. 1991, Preliminary version of the Third
Catalog of Nearby Stars, on: Brotzmann, L.E., Gesser, S.E. (eds.) The
Astronomical Data Center CD-ROM: Selected Astronomical Catalogs,
Vol. I;, NASA/Astronomical Data Center, Goddard Space Flight Center,
Greenbelt, MD 

\bibitem[\protect\citeauthoryear{Grimm et al.}{2002}]{ggs02} Grimm,
  H.-J., Gilfanov, M., \& Sunyaev, R. 2002, A\&A, 391, 923

\bibitem[\protect\citeauthoryear{G\"{u}del}{2004}]{g04} G\"{u}del,
M. 2004, A\&ARv, 12, 71

\bibitem[\protect\citeauthoryear{Hands et al.}{2004}]{hww+04} Hands,
A.D.P., Warwick, R.S., Watson, M.G., \& Helfand, D.J. 2004, MNRAS,
351, 31

\bibitem[\protect\citeauthoryear{Hearty et al.}{2000}]{hns+00} Hearty,
T., Neuh\"{a}user, R., Stelzer, B., Fern\'{a}ndez, M., Alcal\'{a},
J.M., Covino, E., \& Hambaryan, V. 2000, A\&A, 353, 1044

\bibitem[\protect\citeauthoryear{Heinke et al.}{2005}]{hge+05} Heinke,
C.O., Grindlay, J.E., Edmonds, P.D., Cohn, H.N., Lugger, P.M., Camilo,
F., Bogdanov, S., \& Freire, P.C. 2005, ApJ, 625, 796

\bibitem[\protect\citeauthoryear{Hessman}{1998}]{h98} Hessman,
F.V. 1998, A\&AS, 72, 515

\bibitem[\protect\citeauthoryear{H\"{u}nsch et al.}{1999}]{hss+99}
H\"{u}nsch, M., Schmitt, J.H.M.M., Sterzik, M.F., \& Voges, W. 1999,
A\&AS, 135, 319 (H99)

\bibitem[\protect\citeauthoryear{Irwin et al.}{2003}]{iab03}
Irwin, J.A., Athney, A.E., \& Bregman, J.N. 2003, ApJ, 587, 356

\bibitem[\protect\citeauthoryear{Jahrei\ss\ \& Wielen}{1997}]{jw97}
Jahrei\ss, H., \& Wielen, R. 1997, ESA SP-402: Hipparcos -- Venice
'97, 402, 675 

\bibitem[\protect\citeauthoryear{Karata\c{s} et al.}{2004}]{kbe+04}
Karata\c{s}, Y., Bilir, S., Eker, Z., \& Demircan, O. 2004, MNRAS,
349, 1069

\bibitem[\protect\citeauthoryear{Littlefair et al.}{2001}]{ldm01}
Littlefair, S.P., Dhillon, V.S., \& Marsh, T.R. 2001, MNRAS, 327, 669

\bibitem[\protect\citeauthoryear{Makarov}{2003}]{m03} Makarov,
V. 2003, ApJ, 126, 1996 (M03)
	
\bibitem[\protect\citeauthoryear{Malhotra et al.}{1996}]{msr+96} 
Malhotra, S., Spergel, D.N., Rhoads, J.E., \& Li, J. 1996, ApJ, 473, 687

\bibitem[\protect\citeauthoryear{Matsumoto et al.}{1997}]{mka+97}
Matsumoto, H., Koyama, K., Awaki, H., Tsuru, T., Loewenstein, M.,\&
Matsushita, K. 1997, ApJ, 482, 133

\bibitem[\protect\citeauthoryear{McArthur et al.}{2001}]{mbl+01}
McArthur, B.E., Benedict, G.F., Lee, J., van Altena, W.F., Slesnick,
C.L., Rhee, J. et al. 2001, ApJ, 560, 907

\bibitem[\protect\citeauthoryear{Mukai \& Shiokawa}{1993}]{ms93}
Mukai, K., \& Shiokawa, K. 1993, ApJ, 418, 863

\bibitem[\protect\citeauthoryear{Muno et al.}{2004}]{mbb+04} Muno,
M.P., Baganoff, F.K., Bautz, M.W., Feigelson, E.D., Garmire, G.P.,
Morris, M.R., Park, S., Ricker, G.R., \& Townsley, L.K. 2004, ApJ,
613, 326

\bibitem[\protect\citeauthoryear{Ottmann \& Schmitt}{1992}]{os92}
Ottmann, R., \& Schmitt, J.H.M.M. 1992, A\&A, 256, 421

\bibitem[\protect\citeauthoryear{Patterson}{1984}]{p84} Patterson,
J. 1984, ApJS, 54, 443

\bibitem[\protect\citeauthoryear{Patterson}{1994}]{p94} Patterson,
J. 1994, PASP, 106, 209

\bibitem[\protect\citeauthoryear{Revnivtsev et al.}{2004}]{rsj+04} Revnivtsev,
M., Sazonov, S., Jahoda, K., \& Gilfanov, M.R. 2004, A\&A, 418, 927 (R04)

\bibitem[\protect\citeauthoryear{Revnivtsev et al.}{2005}]{r+05} Revnivtsev,
M., Sazonov, S., et al. submitted to A\&A

\bibitem[\protect\citeauthoryear{Robin et al.}{2003}]{rrd+03} Robin,
A.C., Reyl\'{e}, C., Derri\`{e}re, \& Picaud, S. 2003, A\&A, 409, 523

\bibitem[\protect\citeauthoryear{Samus et al.}{2004}]{sd+04} Samus,
N.N., Durlevich, O.V., et al. 2004, Combined General Catalogue of
Variable Stars, VizieR On-line Data Catalog: II/250.

\bibitem[\protect\citeauthoryear{Sazonov \& Revnivtsev}{2004}]{sr04} Sazonov,
S.Y., \& Revnivtsev, M.G. 2004, A\&A, 423, 469

\bibitem[\protect\citeauthoryear{Schmidt}{1968}]{s68} Schmidt,
M. 1968, ApJ, 151, 393 

\bibitem[\protect\citeauthoryear{Schmitt et al.}{1990}]{scs+90}
Schmitt, J.H.M.M., Collura, A., Sciortino, S., Vaiana, G.S., Harnden,
F.R., \& Rosner, R. 1990, ApJ, 365, 704

\bibitem[\protect\citeauthoryear{Schwope et al.}{2000}]{shl+00}
Schwope, A.D., Hasinger, G., \& Lehmann, I. 2000, AN, 321, 1

\bibitem[\protect\citeauthoryear{Schwope et al.}{2002}]{sbb+02}
Schwope, A.D., Brunner, H., Buckley, D., Greiner, J., Heyden, K.v.d.,
Neizvestny, S., Potter, S., \& Schwarz, R. 2002, A\&A, 396, 895

\bibitem[\protect\citeauthoryear{Silber et al.}{1994}]{srh+94} Silber,
A.D., Remillard, R.A., Horne, K., \& Bradt, H.V. 1994, ApJ, 424, 955

\bibitem[\protect\citeauthoryear{Singh et al.}{1996}]{sdw96} Singh,
K.P., Drake, S.A., \& White, N.E. 1996, AJ, 111, 2415

\bibitem[\protect\citeauthoryear{Smith \& Dhillon}{1998}]{sd98} Smith,
D.A., \& Dhillon, V.S. 1998, MNRAS, 301, 767

\bibitem[\protect\citeauthoryear{Sokoloski \& Kenyon}{2003}]{sk03}
Sokoloski, J.L., \& Kenyon, S.J. 2003, ApJ, 584, 1021

\bibitem[\protect\citeauthoryear{Strassmeier et al.}{1993}]{shf+93}
Strassmeier, K.G., Hall, D.S., Fekel, F.C., \& Scheck, M. 1993, A\&AS,
100, 173

\bibitem[\protect\citeauthoryear{Sugizaki et al.}{2001}]{smk+01}
Sugizaki, M., Mitsuda, K., Kaneda, H., Matsuzaki, K., Yamauichi, S.,
\& Koyama, K. 2001, ApJS, 134, 77

\bibitem[\protect\citeauthoryear{Suleimanov et al.}{2005}]{srr05}
Suleimanov, V., Revnivtsev, M., \& Ritter, H. 2005, A\&A, 435, 191

\bibitem[\protect\citeauthoryear{Tinney et al.}{1993}]{trm93} Tinney, 
C.G., Reid, I.N., \& Mould, J.R. 1993, ApJ, 414, 254

\bibitem[\protect\citeauthoryear{Thorstensen}{2003}]{t03} Thorstensen,
J.R. 2003, AJ, 126, 3017

\bibitem[\protect\citeauthoryear{Tovmassian}{1998}]{tgk+98}
Tovmassian, G.H., Greiner, J., Kroll, P., Szkody, P., Mason, P.A.,
Zickgraf, F.-J., et al. 1998, A\&A, 335, 227

\bibitem[\protect\citeauthoryear{Verbunt \& Lewin}{2005}]{wl05}
Verbunt, F., \& Lewin, W.H.G. 2005, to appear in "Compact Stellar
X-ray Sources", eds. W.H.G. Lewin and M. van der Klis, Cambridge
University Press; astro-ph/0404136

\bibitem[\protect\citeauthoryear{Voges et al.}{1999}]{vab+99} Voges,
W., Aschenbach, B., Boller, Th., Br\"{a}uninger, H., Briel, U.,
Burkert, W., Dennerl, K. et al. 1999, A\&A, 349, 389

\bibitem[\protect\citeauthoryear{Warner}{1995}]{w95} Warner,
B. Cataclysmic variable stars. Cambridge Astrophysics Series,
Cambridge University Press 1995
	
\bibitem[\protect\citeauthoryear{Warwick et al.}{1985}]{wtw+85}
Warwick, R.S., Turner, M.J.L., Watson, M.G., \& Willingale, R. 1985,
Natur, 317, 218

\bibitem[\protect\citeauthoryear{Worrall et al.}{1982}]{wmb+82} Worrall, D.M.,
Marshall, F.E., Boldt, E.A., \& Swank, J.H. 1982, ApJ, 255, 111

\bibitem[\protect\citeauthoryear{Worrall \& Marshall}{1983}]{wm83}
Worrall, D.M., \& Marshall, F.E. 1983, ApJ, 267, 691

\end{thebibliography}
\end{document}